\documentclass{article}

\usepackage{PRIMEarxiv}

\usepackage[utf8]{inputenc} 
\usepackage[T1]{fontenc}    
\usepackage{hyperref}       
\usepackage{url}            
\usepackage{booktabs}       
\usepackage{amsfonts,amsmath}       
\usepackage{nicefrac}       
\usepackage{microtype}      
\usepackage{lipsum}
\usepackage{fancyhdr}       
\usepackage{graphicx,xcolor}       
\graphicspath{{media/}}     

\title{Local interactions in active matter are reinforced by spatial structure 
}

\author{
  Eighdi Aung\\
  Department of Biomedical Engineering and Mechanics \\
  Virginia Tech \\
  Blacksburg, Virginia, USA \\
  \texttt{eighdiaung@vt.edu} \\
   \And
  Nicole Abaid\\
  Department of Mathematics \\
  Virginia Tech \\
  Blacksburg, Virginia, USA \\
  \texttt{nabaid@vt.edu} \\
   \And
  James E. McClure\\
  National Security Institute \\
  Virginia Tech \\
  Blacksburg, Virginia, USA \\
  \texttt{mcclurej@vt.edu} \\
}

\begin{document}
\maketitle

\begin{abstract}
The flocking of self-propelled particles in heterogeneous environments is relevant to both natural and artificial systems. The Vicsek model is a canonical choice to investigate such systems due to the minimal number of parameters required to define flocking. Prior research on the Vicsek model has investigated the effects of interaction rules, particle speed, and obstacle packing on the flocking behavior, but the effect of interaction radius remains an open question. Unlike obstacle-free domains, the locality of interactions not only affects how quickly the system can become polarized, but also how well the flocks can align or realign after colliding with obstacles. In this letter, we delve into this subtle relationship that exists in the scale of the perception of Vicsek particles in the presence of obstacles. We demonstrate that the presence of obstacles impacts group density, which provides the basis to identify distinct phases for collective behavior. This leads to the counter-intuitive result that obstacles, while generally confounding for macroscopic order, may enable global order even as noise in the system increases.
\end{abstract}

\keywords{Vicsek with Obstacles \and Phases of Flocking \and Active Matter \and Spatial Heterogeneity}

\section{Introduction}
Active matter comprises a broad class of physical systems defined by emergent collective behaviour, ranging from bird flocks and human crowds to the dynamic interactions of bacteria and tissues. In contrast with standard diffusion processes described by Brownian motion, self-propelled particles can align their motion to cover distance more efficiently. The associated scaling behavior of mean square displacement with time is thereby super-diffusive \cite{chate2008review}. This tendency is confounded by the presence of obstacles, since confinement produces a sub-diffusive effect \cite{PhysRevE.80.011109,PhysRevLett.98.250602}. It follows that active matter is subject to competing effects within confined environments. 
How well groups of self-propelled particles can coordinate their motion in an environment constrained by structure is a compelling topic due to the existence of natural \cite{cisneros2006reversal,battersby2015,evangelista2017,ning2023} and artificial \cite{gokce2010,varga2022} systems that perform flocking in the presence of obstacles. The essence of flocking in such environments relies on the interplay of interactions among particles, as well as their response to the constraints the environment imposes on their movement. We demonstrate that the presence of obstacles alters the dominant polarization regimes for active matter, and that these regimes are controlled by the ratio of the length scale for interactions relative to the size of the structures. Obstacles can also lead to the formation of more tightly packed flocks, such that structural heterogeneity within the system is reinforced as a consequence of this interaction. 

The canonical Vicsek model \cite{vicsek95,ginelli_2016review} generates flocking using a minimal amount of parameters to define the dynamics. Many variations of the Vicsek model have been explored \cite{gregoire2004,chate2008review,degond2008,zhang2009,baggaley2015,costanzo2022,chatterjee2023}. In recent years, the behavior of the Vicsek model in environments with obstacles has been examined. Martinez et al. \cite{martinez2018} performed a comprehensive study of such a system by investigating how much flocking is achievable according to the type of collision rules imposed, the particle speed, and the obstacle packing fraction. Rahmani et al. \cite{rahmani2021} extended the study of the Vicsek model with obstacles by using topological interaction rules and showed that, even for metric-free interactions, flocking arises due to spatial heterogeneity  and is density dependent. Furthermore, Codina et al. \cite{codina2022} showed that a single minuscule obstacle can produce flock reversal of Vicsek particles. Most recently, Serna et al. \cite{serna2023} identified the existence of phases in the scale of flocking in a Vicsek system as a function of the spacing between obstacles in the environment. 

To further the understanding of the Vicsek system with obstacles, we focus on the effect of the interaction radius on flocking. The key feature of the Vicsek model is the ability of the system to self-organize into flocks from local interactions. In an obstacle-free domain, the extent to which the system is ``local" would naturally be defined by the particular biological or physical system we are investigating. For example, to model a bird flock, we would use the visual range to define the particles' interaction radius. In the case of low noise in such a system, we can expect that the interaction radius would only affect how quickly the system can reach global polarization. However, the same cannot be said for a Vicsek system with obstacles because, even in the absence of noise, the system will never sustain global polarization. Even more so, the system may have multiple steady-states for which global polarization may occur. Hence, in heterogeneous environments, the locality of interaction impacts the flock's ability to retain alignment or realign after colliding with obstacles. In this letter, we explore this nuanced relationship between the scale of a particle's perception and quenched disorder that exists in a system with obstacles. Specifically, we show the existence of different phases of flocking due to the coupling that exists between the particle's interaction range and the size of the obstacles in the environment. 

\section{Vicsek model with obstacles}

We use the Vicsek model \cite{vicsek95} with $N$ particles in a periodic box of length $L$. The $i^{\mathrm{th}}$ particle at timestep $t$ has position $\textbf{x}_{i}(t) \in [-L/2,L/2]^2$ and velocity $\textbf{v}_{i}(t) \in\mathbb{R}^2$. At each timestep, the particle heads in the direction $\theta_{i}(t+1)$, which is calculated by averaging the heading of neighboring particles within an interaction radius, $R$. This averaging is subject to noise, $\xi$, which is sampled uniformly from $[-\pi,\pi]$ with strength $\eta$. Thus, the heading update of the $i^{\mathrm{th}}$ particle with neighbors indexed $j$ is:
\begin{equation}
    \theta_{i}(t+1) = \langle \textnormal{arg}[\textbf{v}_{j}(t)] \rangle_{R} + \eta\; \xi,
\end{equation}
The particle then updates its velocity and position:
\begin{align}
    &\textbf{v}_{i}(t+1) = \textnormal{cos}[\theta_{i}(t+1)] \textbf{e}_1 + \textnormal{sin}[\theta_{i}(t+1)] \textbf{e}_2,\\
    &\textbf{x}_{i}(t+1) = \textbf{x}_{i}(t) + c\, \textbf{v}_{i}(t+1) ,
\end{align}
where $\textbf{e}_1$ and $\textbf{e}_2$ are unitary basis vectors of $\mathbb{R}^2$, and $c$ is the constant speed of the particle. With the addition of obstacles to the domain, we add the condition that, if the updated position of the particle is inside an obstacle, then the particle performs an avoidance maneuver on heading while remaining fixed for that timestep:
\begin{align}
    &\textbf{v}_{i}(t+1) = \textbf{v}_{i}(t) \times \beta \, \textbf{e}_3,\\
    &\textbf{x}_{i}(t+1) = \textbf{x}_{i}(t),
\end{align}
where $\beta$ is a binomial random choice between $\{-1,1\}$ multiplying the unitary vector $\textbf{e}_3 = \textbf{e}_1 \times \textbf{e}_2$ with $\textbf{v}_{i}(t)$, $\textbf{e}_1$, and $\textbf{e}_2$ nominally extended to three dimensions to perform the cross product. We choose to model obstacle avoidance in this way to allow the particle to make an unbiased choice between turning left or right orthogonally to their previous heading.

\begin{figure}
    \centering
    \includegraphics[width=0.6\columnwidth]{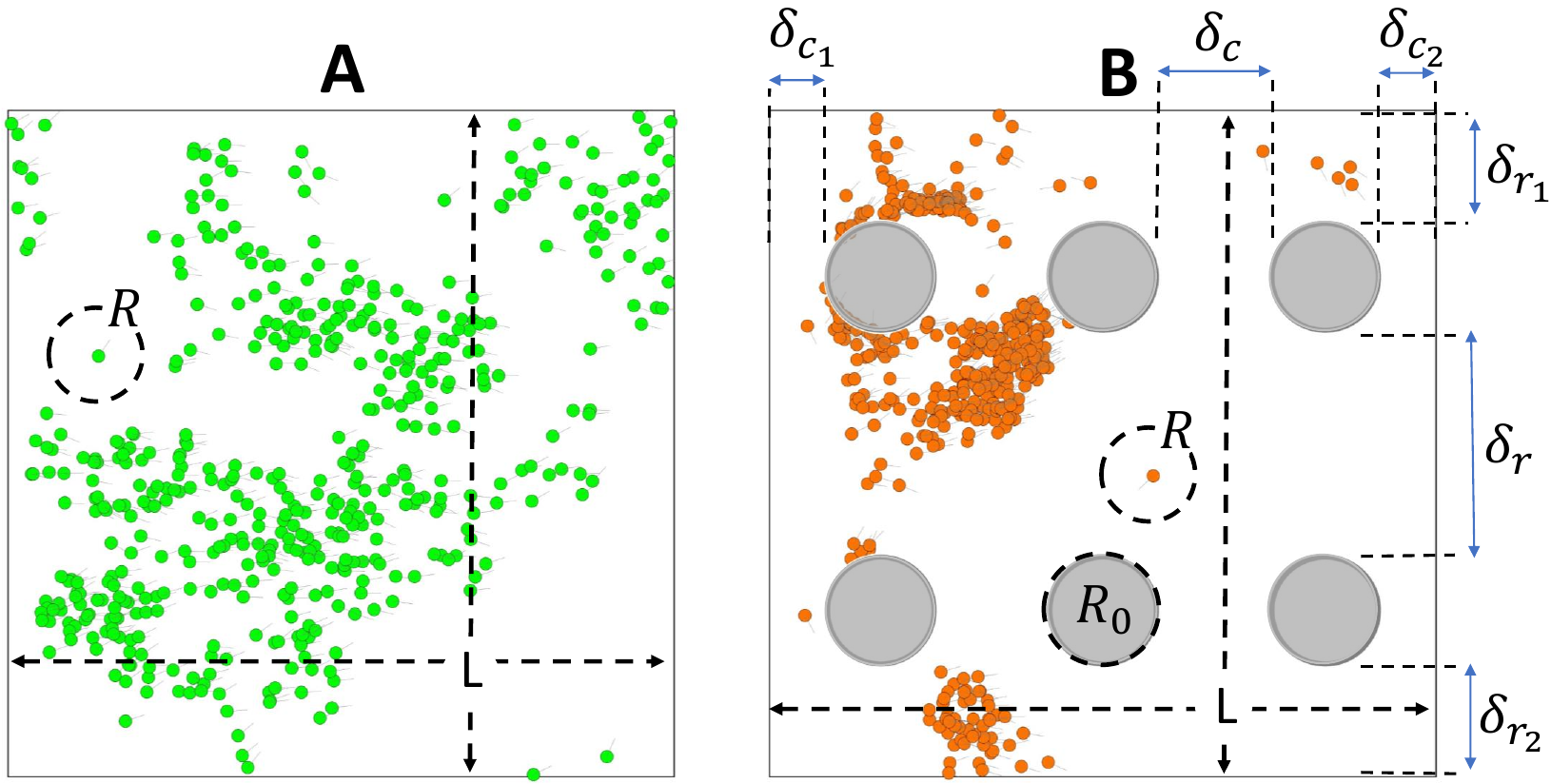}
    \caption{Figures A and B show a snapshot of Vicsek particles without and with obstacles, respectively, with interaction radius $R = 0.28$ and $\eta = 0.12$, confined in periodic boxes of length $L$. The grey circles in figure B indicate obstacles with radius $R_o$. 
    The obstacles are equidistantly arranged with horizontal and vertical gap spacings $\delta_c$ and $\delta_r$. 
    To ensure equal spacing around the periodic boundaries, 
    $\delta_{c_1}+\delta_{c_2}=\delta_{c}$ and $\delta_{r_1}+\delta_{r_2}=\delta_{r}$.}
    \label{fig:domain}
\end{figure}

Figure \ref{fig:domain}A shows a snapshot of a canonical Vicsek system. As shown in Figure \ref{fig:domain}B, we designed the domain to have two rows of three obstacles. We spaced the obstacles equidistantly from the boundaries horizontally and vertically such that, $\delta_c = \delta_{c_1} + \delta_{c_2}$ and $\delta_r = \delta_{r_1} + \delta_{r_2}$, respectively. For this study, the periodic box has length $L = 5$ and the  obstacles have $R_o = 0.4$, making $\delta_r = 1.70$ and $\delta_c = 0.87$. The density, $\rho$, is the ratio of the total number of particles, $N$, to the unobstructed area; we set $\rho=20$ which means $N=500$ without obstacles and $N=440$ with obstacles. 
We fix the speed to be $c=0.05$, so that particles need on the order of 100 timesteps to traverse the domain. Hence, we only vary $R$ and $\eta$. 

For a Vicsek system without obstacles, the interaction radius is naturally bounded by $L$. On the other hand, with obstacles, the interaction between particles is influenced by obstacle size, which has $R_o<<L$. We propose a characteristic radius, $r$, to be the ratio of the interaction radius to the obstacle size. For this study, we focus on $\eta \in [0,0.5]$ and $r = R/R_o \in [0.1,6.25]
$. For each choice of noise and interaction radius, we run 100 replicate simulations of 10,000 timesteps with random initial conditions.

\section{Polarization and coverage as order parameters}

In the Vicsek model, the inter-particle interaction is defined as heading alignment. Hence, the natural choice for an order parameter is polarization, which is the magnitude of the sum of the particles' velocities normalized by particle number and constant speed. We compute the steady-state mean polarization over $T$ timesteps:
\begin{equation}
    \varphi = \left< \frac{1}{cN} \left  \| \sum^{N}_{i=1}  \textbf{v}_{i}(t) \right \| \right>_T.
\end{equation}
We define steady-state to be when there is no significant change in the mean and variance of the order parameter, which is reached after $2000$ timesteps, making $T = 8000$.

\begin{figure}
    \centering
    \includegraphics[width=0.6\columnwidth]{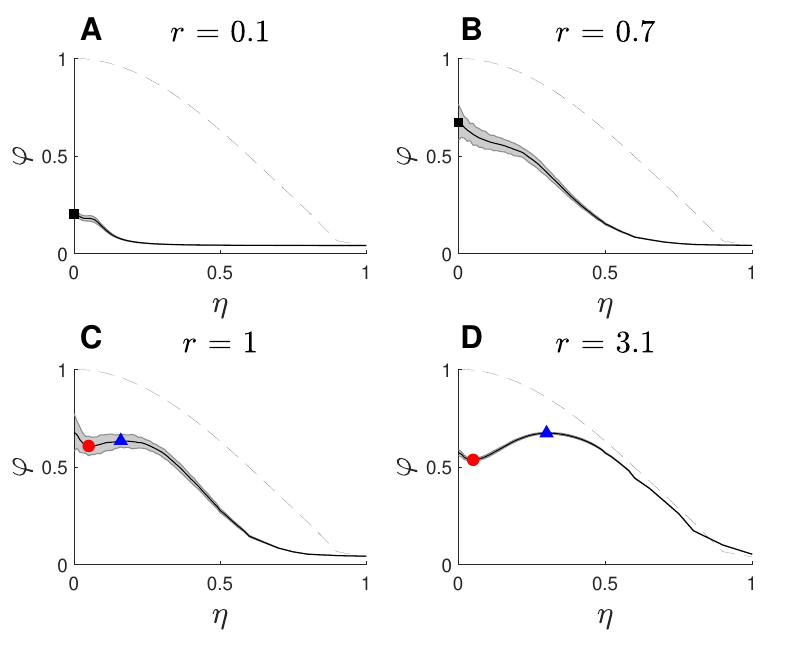}
    \caption{Mean polarization over $T = 8000$ timesteps, as a function of noise, $\eta$, for example cases of the characteristic radius, $r$. Curves indicate the mean polarization over 100 Monte Carlo replicates and shaded regions show $\pm$ one standard deviation. The black square markers indicate the location of global maxima when the curves do not have critical points. This behavior is observed in figures A and B, which have interaction radii less than the radius of the obstacles, i.e., $r < 1$. When $r \geq 1$, local extrema appear. The local minimum and maximum are marked by red circles and blue triangles, respectively, shown in figures C and D. The dashed lines show the polarization curves obtained using the same value of $r$ for a Vicsek system without obstacles. Even though $R_0$ is not meaningful to this system, $r$ is shown as $R$ scaled by a constant $R_0$ to allow direct comparison when obstacles are introduced.}
    \label{fig:pol_curves}
\end{figure}

We first explore how polarization changes with noise and interaction radius when obstacles are present. We expect that polarization will be strictly less than one (i.e, $\varphi < 1$), even in the absence of noise (i.e, $\eta = 0$), since particles are sure to encounter obstacles and disperse frequently. Furthermore, recent literature \cite{martinez2018,mcclure2022} has shown that, at low noises, polarization as a function of noise exhibits non-monotonicity in contrast with the Vicsek model without obstacles. The non-monotonic behavior is driven by the particles' collisions with obstacles and noise. Figure \ref{fig:pol_curves} shows example cases of polarization as a function of noise for increasing values of interaction radius. The top plots, for which $R<R_o$, do not have the characteristic concavity of the polarization plots for the Vicsek model without obstacles. They also attain much lower maximum values of polarization by comparison. Interestingly, the bottom plots, for which $R\geq R_o$, show the appearance of local extrema of the polarization curves. The order in which the local minimum and maximum points appear is directly tied to noise. At low noise, the system consists of tightly packed flocks, as opposed to a noisier system where the flocks are dispersed. The dispersal of flocks due to noise compensates for the reduction in polarization due to collisions with obstacles, consistent with \cite{chepizhko2013}. Additionally, we observe that increasing interaction radius widens the spacing between the two critical points. 

\begin{figure*}
\includegraphics[width=\columnwidth]{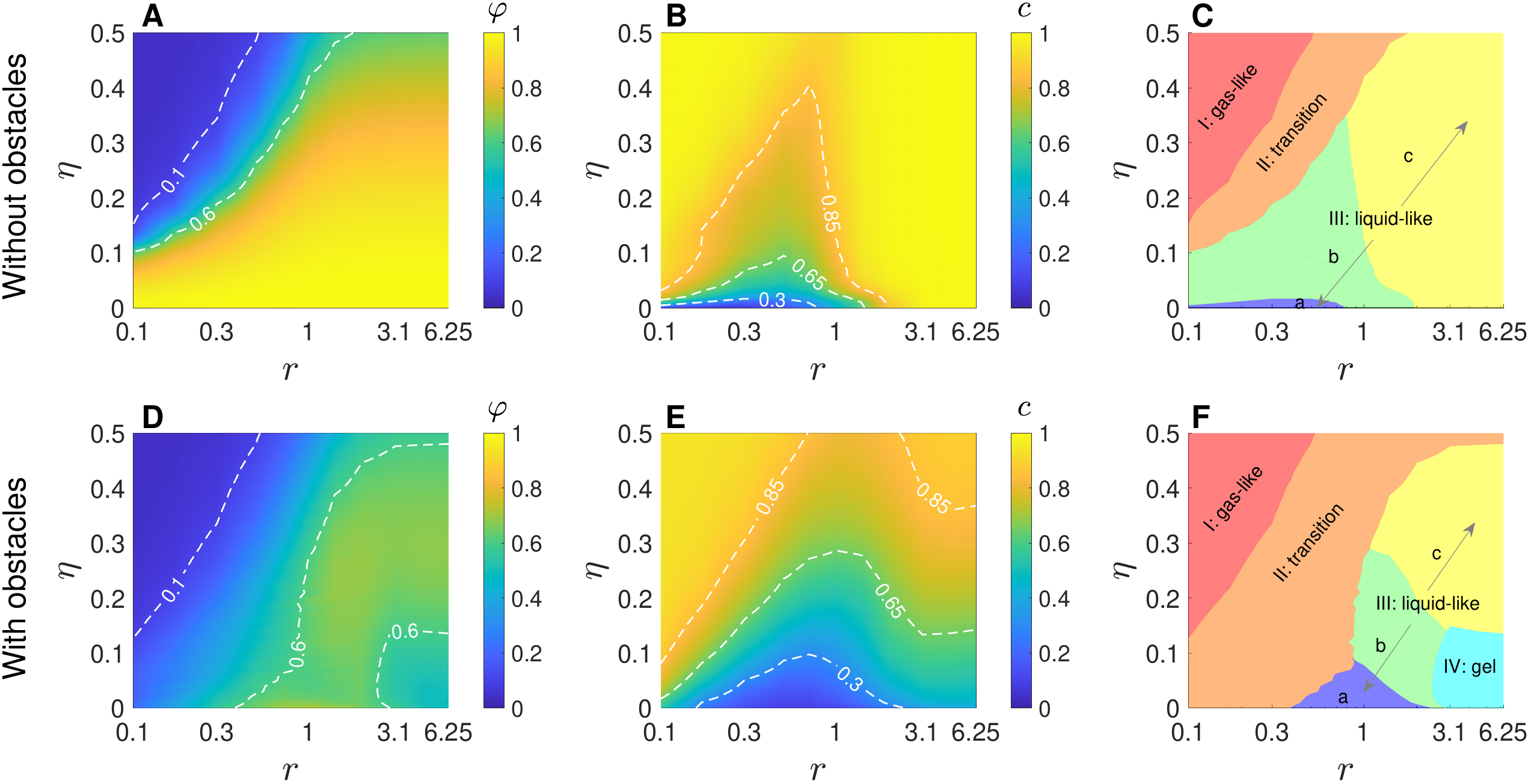}
\caption{Figures A, B and C show polarization, coverage, and the resulting phase diagram in the parameter space of $\eta$ and $r$, respectively, for the Vicsek system without obstacles. Figures D, E and F show the same quantities for the Vicsek system with obstacles. The horizontal axes for all the figures are use a logarithmic scale. We observed three main regimes of polarization which we identify from drawing contours of $\varphi = \{0.1,0.6\}$ for figures A and D,  and four regimes of coverage separated by contours of $c = \{0.3,0.65,0.85\} $ for figures B and E. The contours of polarization and coverage are used to identify phases shown in figures C and F. Red (I) phases are weakly polarized and no general flocking motion can be observed, hence, gas-like. Flocks start to form in the orange (II) transition phases. Highly polarized flocks are observed in indigo (III a), green (III b) and yellow (III c) phases. The blue (IV) phase, only observed when obstacles are present, shows the highest degree of persistence for clusters.}

\label{fig:heatmaps}
\end{figure*}

Figures \ref{fig:heatmaps}A and D show heatmaps of $\varphi$, denoted on the coloraxis, as a function of $r$ and $\eta$. We observe two general regimes of polarization in both figures, however we do not see high polarization values ($\varphi > 0.6$) in the presence of obstacles. We define threshold polarization values, $\varphi = \{0.1,0.6\}$, for the Vicsek system with obstacles (Figure \ref{fig:heatmaps}D) to separate low, intermediate, and high polarization regimes. This same threshold is used for Figure \ref{fig:heatmaps}A for a fair comparison between the two systems.
For all values of $\eta$, horizontal slices in Figure \ref{fig:heatmaps}A show a monotonic increase, as expected of a Vicsek system without obstacles. We see similar behavior, although only lower polarization values are attained, when obstacles are present and $\eta>0.1$ in Figure \ref{fig:heatmaps}D.
In stark contrast, for $\eta \leq 0.1$ with obstacles, we observe the reappearance of the intermediate regime in the lower-right corner, meaning that polarization attains a maximum value for fixed noise at some intermediate value of $r$. This feature is directly tied to the existence of local critical points in the polarization curves when the obstacles are present. As seen from the polarization curves, Figure \ref{fig:pol_curves}C and D, the existence of local maximum and minimum polarizations for values of $r \geq 1$ depends solely on $\eta$. Therefore, we require an additional order parameter that can decouple the effects of $r$ and $\eta$. From observing the system dynamics, we determine that spatial coverage is a reasonable and simple choice to parameterize the effect of $\eta$. We quantify coverage, $c$, as the steady-state time averaged fraction of area occupied by particles:
\begin{equation}
    c = \left< \frac{b_o(t)}{b} \right>_T,
\end{equation}
where we discretize the domain into $b=100$ total bins and $b_o$ is the number of bins occupied by particles at time $t$. Figures \ref{fig:heatmaps}B and D are the heatmaps of coverage for systems without and with obstacles, respectively. For both systems, there are four general regimes separated by the three contours $c \in \{0.3,0.65,0.85\}$. 
The sensitivity of coverage to noise is strikingly evident from vertical stacking of prominent regimes in the parameter space.

\section{Phases of flocking with and without obstacles}

The regimes identified from the polarization and coverage heatmaps are used to define dynamical phases that the Vicsek system exhibits in parameter space  (see Supplemental Material for a movie of examples). Figure \ref{fig:heatmaps}C shows the phases observed in the Vicsek model. We identify the red (I) phase where the system is weakly polarized and the particles are moving in different directions. In this phase, the system is qualitatively  gaseous. Either increasing interaction radius or decreasing noise brings the system to a transition phase, colored orange (II), where it starts to form local order but does not have consistently cohesive flocks due to retaining randomness from the red phase. Reducing $\eta$ further brings the system to the three phases shown in yellow (III c), green (III b), and indigo (III a). In these three phases, the system is highly polarized. In the yellow phase, the system is most dispersed and covers most of the available space. At the other extreme, the indigo phase consists of small clusters of highly dense flocks. The existence of the indigo phase even in the lowest case of $r$ shows that a Vicsek system can reach consensus as long as it is allowed to evolve over enough time. This suggests that the ``locality" of interactions in a Vicsek system without obstacles is only bounded below by $r=0$. Additionally, we identify a green phase between yellow and indigo, where the particles are moderately dispersed and may appear most like natural flocks. Qualitatively, we identify the  yellow, green, and indigo phases to be liquid-like, where macroscopic flow in the movement of the flock is observed.

We identify similar phases for a Vicsek system with obstacles, though the presence of obstacles noticeably affects the phase diagram, see Figure \ref{fig:heatmaps}F. The first noticeable change is in the growth of the orange phase, overtaking the yellow, green, and indigo phases of high polarization. This effect is intuitive since we already know that obstacles introduce more disorder in the system. Interestingly, the indigo phase no longer starts at bottom left corner, indicating that there is a non-zero lower bound for the interaction radius where macroscopic flocking in the presence of obstacles can be achieved, even in the absence of noise. Additionally, the indigo region also extends to higher noises than in the case without obstacles, showing that the obstacles can also produce the counter-intuitive effect of promoting dense clusters. The green and yellow regions now start at $r \approx 1$, which indicates that dispersed flocking phases are achieved when $R\approx R_o$. 
The most interesting change in the phase diagram with obstacles is the appearance of the blue (IV) phase, which ties to the non-monotonic behavior of the polarization curves. In this phase, the system quickly reaches alignment such that the particles move in synchrony in a rigidly locked state. When any part of the group encounters an obstacle in this interlocked state, the entire group responds, causing particles to experience ``virtual" obstacles, even when they are in open space. In this phase, the flocks behave like a viscous liquid which temporarily ``sticks" to the obstacles. The sticking is directly caused by the interaction term dominating the obstacle maneuver consistently over time for dispersed flocks. We note that this interlocked flocking phase is latent in Figure \ref{fig:heatmaps}C, but is indistinguishable from the rest of the yellow phase due to the absence of obstacles.

\section{Conclusion}
In conclusion, we observed that the interaction radius profoundly affects the behavior of the Vicsek system, and even more so, that such an effect is most prevalent in the presence of obstacles. Analogous to a thermodynamic system, the noise acts like temperature and the interaction radius acts like intermolecular bonding for the flocks. As such, we generally find that increasing noise (i.e., moving rightwards) in the phase diagrams leads to states of higher entropy, and increasing interaction radius (i.e., moving upwards) in the phase diagrams leads to states of lower entropy. The latter trend is more striking in the presence of obstacles, which implies that the ``locality" in the interactions matters more when space is constrained. Additionally, we see that a non-zero minimal interaction radius is required to achieve substantial flocking in the system. Most interestingly, the obstacles may promote order within the system. This is shown by the growth of the liquid-like phases in parameter space, as well as the ability of the system to reach a further ``cooled" gel-like phase that manifests in the presence of obstacles with large interaction radius. 

\section*{Acknowledgments}
This work was supported by the National Science Foundation under award 1751498. The authors are grateful to Advanced Research Computing at Virginia Tech for computing resources.

\bibliographystyle{unsrt}  
\bibliography{references}

\end{document}